\documentclass{article}
\usepackage{graphicx}

\usepackage[margin=.75in]{geometry}

\begin{document}

\newtheorem{theorem}{Theorem}[section]
\newcommand{\EQ}{Eq.~}
\newcommand{\EQS}{Eqs.~}
\newcommand{\FIG}{Fig.~}
\newcommand{\FIGS}{Figs.~}
\newcommand{\TAB}{Tab.~}
\newcommand{\SEC}{Sec.~}
\newcommand{\SECS}{Secs.~}

\begin{center}
{\Large
Filtered interspike interval encoding by class II neurons}\\
\bigskip
Naoki Masuda, Kazuyuki Aihara\\
\bigskip
Department of Complexity Science and Engineering,\\
Graduate School of Frontier Sciences, the University of Tokyo,\\
7--3--1, Bunkyo-ku, Tokyo, 113--8656, Japan\\
\bigskip

\end{center}

\begin{abstract}
Dynamics of class II neurons, firing frequencies of which are strongly regulated by the inherent neuronal property, have been extensively studied since the formulation of the Hodgkin--Huxley model in 1952. However, how class II neurons process stimulus information and what kind of external information and internal structure firing patterns of neurons represent are vaguely understood in contrast to firing rate coding by class I neurons. Here we show that the FitzHugh--Nagumo class II neuron simultaneously filters inputs based on the input frequency and represent the signal strength by interspike intervals. In this sense, the class II neuron works as an AM processor that passes the information on the carrier and on the temporal waveform of signals.
\end{abstract}

\bigskip



\bigskip

Oscillation with a dominant frequency is often found in neural systems
both {\it in vitro} and {\it in vivo} in spite that the inputs to neurons
and the neurons themselves are quite noisy in general.
For example,
inferior olive neurons of the guinea pig regularly fire with
frequencies in the theta band (3--10 Hz) \cite{Benardo,Llinas86}.
Local field potentials oscillate
in the gamma band (mainly 30--40 Hz) in the rat hippocampus
\cite{Bragin} and the cat cortex \cite{Steriade}, and spiking
with almost regular interspike intervals corresponding to 
this frequency has been found as well
\cite{Bragin,Gutfreund,Traub96}.
These regular oscillations may be indicative of such cortical 
functions as feature binding \cite{Steriade,Gray},
increased alertness of animals \cite{Steriade},
and cognition \cite{Llinas93}.
However, it is still in dispute what kind of information oscillations
themselves encode, how the information is exchanged via oscillatory
signals, and how neural networks treat them to yield sophisticated
brain functions.  Our primary concern is to examine the neuronal
mechanism of dynamical information processing with oscillations.

To cope with this problem, let us return to the mechanism of
oscillations. Biologically, regular oscillations in neural networks can
be generated by intrinsic cellular mechanisms
\cite{Gutfreund,Aihara82}, rhythmic inputs such as auditory
stimuli \cite{Lu01}, and/or
mutual interaction among neurons.  From a theoretical viewpoint, model neurons are
classified into two classes in terms of excitability 
\cite{Izhikevich00}. As the bias applied to the neuron is increased from the
subthreshold level, class I neurons begin firing with an
infinitesimally low frequency typically via the saddle-node
bifurcation.  The firing frequency is sensitive to the bias strength,
and regular oscillations are unlikely to occur with highly fluctuating
inputs, although pioneering work showed that noise in such a
dynamical system can induce some degree of temporal regularity
\cite{Sigeti}.  On the other hand, the firing frequency of a
class II neuron jumps from zero to a finite value with increasing the
bias usually via the Hopf bifurcation.  The frequency is mainly
determined by the internal dynamics and therefore less sensitive to
the bias.  The class II property can serve to generating regular
oscillations even under noisy conditions.


From a more computational or operational viewpoint, class I neurons
have attracted much more attention. A class I neuron such as an
integrate-and-fire neuron works as an
integrator or a coincidence detector of inputs when the membrane time
constant is large or small, respectively.  In the former case,
interspike intervals, or equivalently instantaneous firing rates, encode the
strength of temporally changing inputs
\cite{Sauer94,Racicot,Masuda_ISI}.  When coupled, class I
neurons are capable of information transmission via the synfire chain
\cite{Abeles91,Diesmann} or the population rate coding
\cite{Shadlen98,Mar}. Feedforward networks of class I neurons
also show switching
between the inter-synchronization-interval code, namely, a code
representing instantaneous firing rates by the inverse of
intervals of synchronous firing, and the rate code \cite{Masuda_IEI,Masuda_DUAL,Vanrossum}.
These investigations have been motivated partly because class I neurons, rather
than II, are better approximation to the cortical neurons responsible
for many central functions in the brain.

In contrast, how dynamics of oscillatory class II neurons
contributes to information processing is still unclear.  Actually,
class II neurons show interesting phenomena such as the coherence
resonance \cite{Longtin97,Pikovsky} in which a moderate amount
of noise reinforces periodic firing with the intrinsic frequency of
the neuron. However, regular oscillations themselves do not seem to
encode external stimuli except that an assembly of synchronously or
coherently oscillating neurons may be associated to a particular
feature \cite{Gray}.
This caution also applies to the abundant results on purely regular
oscillations with constant inputs.  Furthermore, the interspike
interval coding mentioned above is in most cases difficult for class
II neurons because the firing frequency does not strongly depend on
the input strength \cite{Racicot,Masuda_ISI}. One of
the few remarkable observations in the functional respect is that
class II neurons work as resonators that tend to fire when the input
frequency is resonant with the inherent frequency of the neuron
\cite{Izhikevich00}. This fact is also supported by
experimental results \cite{Gutfreund,Hutcheon00}.
Izhikevich has demonstrated some possible applications of class II
neurons as resonators to, for example, a filter of spike doublets and
its multiplexing in which doublets from various sources with different
internal frequencies are transmitted through a common channel
\cite{Izhikevich00,Izhikevich01}. However, these mechanisms require
that each doublet is sufficiently apart from each other in time
without being interfered by other spikes.  This seems unrealistic in real
situations where the neurons are subject to more complicated inputs
such as triplets, persistent oscillations with possibly varying
amplitude and frequency, and noise.

In this Letter, we examine how class II neurons filter and express
complex inputs by evaluating the performance of the interspike
interval coding depending on the input frequency. We first note that if
a neuron responded to the inputs in an all-or-none manner in terms of
the firing frequency, that is, if the neuron fired exactly 
with its inherent frequency
with a suprathreshold input and stayed totally quiescent otherwise, it
would convey just one bit of information about temporally changing inputs.  The
all-or-none behavior is actually not the case for either seemingly
class II biological
neurons 
\cite{Gutfreund,Traub96} or class II model neurons such as the
Hodgkin--Huxley and FitzHugh--Nagumo (FHN) neurons \cite{Izhikevich00}.
Even though the firing frequency jumps from zero to a finite value with
an increasing bias, the frequency gradually increases to some
extent in response to
further increase in the bias. The frequency modulation
has been experimentally found 
in the theta rhythm of inferior olive neurons
under current or pharmacological application \cite{Benardo,Llinas86} and
in the hilar gamma rhythm under cortex
lesion \cite{Bragin}.  Such
modulated outputs are indicative of oscillations whose amplitudes are
temporally changing, which are considered to carry information on stimulation
or behavioral conditions in the case of theta \cite{Benardo,Llinas86} and
gamma \cite{Bragin,Traub96} oscillations. Consequently, we expect that a
class II neuron encodes the inputs in the form of interspike intervals
or instantaneous firing rates only
if the input lies in the limited frequency band. In the following, we
simultaneously deal with two functional properties of a neuron,
that is, possibility
of the interspike interval coding with temporally changing inputs
and possibility of input filtering determined by the class of excitability.

We use a FHN neuron \cite{Fitzhugh,Nagumo62}
whose dynamics is defined by
\begin{eqnarray*}
\alpha \frac{dv}{dt} &=& -v (v-0.5) (v-1) - w + S(t),\\ 
\frac{dw}{dt} &=& v - w - 0.15,
\end{eqnarray*}
where $\alpha \ll 1$, $S(t)$ is the external input, $v$ is the fast
variable representing the membrane potential,
and $w$ is the slow variable. The threshold for firing is set at $v=0.7$.
The value of $\alpha$ specifies the inherent frequency of the neuron.

We start by applying a chaotic input that
models complex but non-random
external stimuli such as sounds and odors.
The input is defined to be
$S(t) = 0.23 + 0.0075 x$
where $x$ is generated from 
the R\"{o}ssler 
equations \cite{Sauer94,Racicot,Masuda_ISI}
represented by
$\dot{x} = a (-y-z)$, $\dot{y} = a (x+0.36y)$,
$\dot{z} = a (0.4+z(x-4.5))$.
The R\"{o}ssler system is chaotic but close to periodic in the sense
that it has a dominant frequency proportional to $a$, which we vary
to examine the filtering property of the FHN neuron.
For class I neurons such as integrate-and-fire
neurons, interspike intervals can carry deterministic information 
when the neurons receive
continuous chaotic inputs like $S(t)$ \cite{Sauer94,Racicot}
or chaotic spike inputs from other neurons \cite{Masuda_ISI}. 
However, the interspike interval coding
is difficult for FHN neurons because 
they do not operate as efficient amplitude-to-frequency converters
\cite{Masuda_ISI}. The determinism of the inputs is passed through only when 
the input strength is skillfully
chosen or it changes very slowly \cite{Racicot}.
Here we focus on the influence of the input frequency.

The determinism of the output is measured
by the deterministic prediction error of interspike intervals.
With $T_k$ denoting the time of the $k$th firing,
the interspike intervals are given by $t_k = T_k - T_{k-1}$.
Then the time series $\{t_k\}$ is predicted by
the local prediction algorithm widely used for chaotic time
series \cite{Sauer94,Racicot,Masuda_ISI}. To this end,
we transform $\{t_k\}$ into
the $d$-dimensional state points $\{ (t_k,t_{k-1},\ldots,t_{k-d+1}) \}$.
To predict $t_{k_0+h}$ at time $t_{k_0}$,
we search $l_0$ nearest neighbors
$(t_{k_l},t_{k_l-1},\ldots,t_{k_l-d+1})$,
$1\le l\le l_0$
to the point $(t_{k_0},t_{k_0-1},\ldots,t_{k_0-d+1})$.
Then $t_{k_0+h}$ is predicted by
$\hat{t}_{k_0+h} = \sum^{l_0}_{l=1} t_{k_l+h} / l_0.$
The performance of prediction is evaluated by the normalized
prediction error:
\begin{equation}
NPE(h) = \left<(\hat{t}_{k_0+h} - t_{k_0+h})^2\right>^{1/2} 
\bigg/ \left<(m - t_{k_0+h})^2\right>^{1/2},
\label{eq:npe}
\end{equation}
where $\left<\cdot\right>$ denotes the average over the whole time series,
and $m$ is the mean of $\{ t_k\}$.
The denominator of \EQ\ref{eq:npe} is the expected
error for the prediction by the mere average.

\begin{figure}
\begin{center}
\includegraphics[height=1.5in]{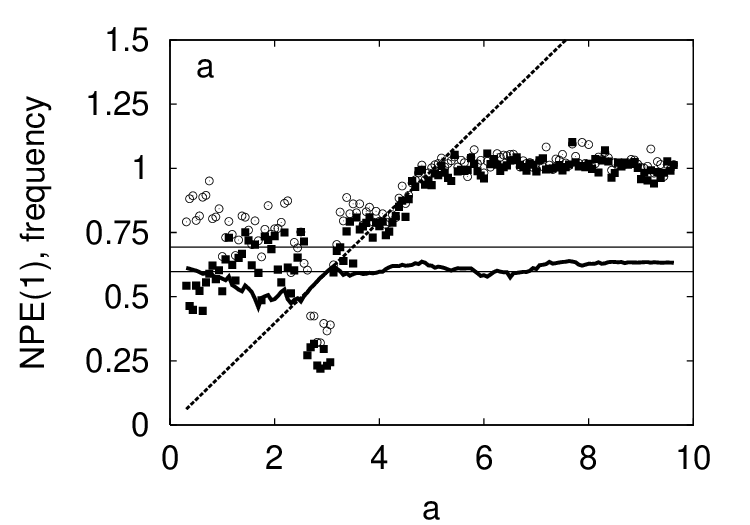}
\includegraphics[height=1.5in]{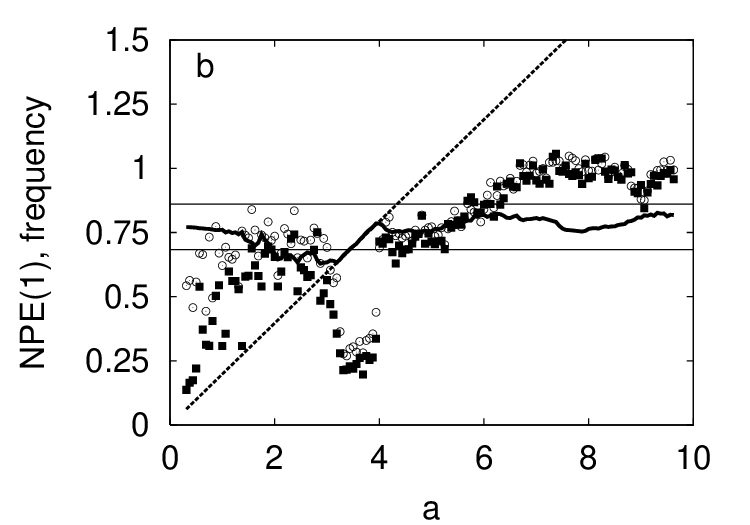}
\includegraphics[height=1.5in]{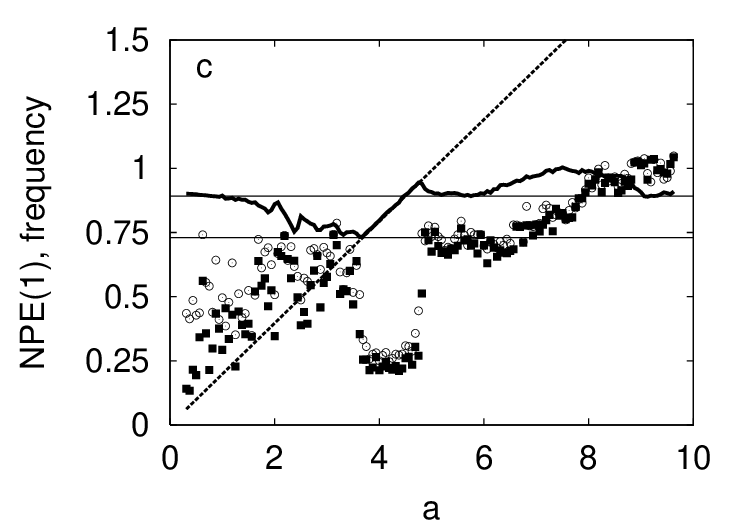}
\caption{Performance of interspike interval coding of a FHN neuron
measured by $NPE(1)$ in the noiseless case (solid squares) and the
noisy case (open circles). The neuron is driven by the R\"{o}ssler
inputs, and $a$ is varied. The frequency of inputs (dotted line), the
actual firing frequency of the FHN neurons (thick solid line), and the
inherent firing frequency range of the neurons with constant inputs
(thin solid lines) are also shown.  (a) $\alpha=0.05$, (b)
$\alpha=0.03$, and (c) $\alpha=0.02$.}
\label{fig:r}
\end{center}
\end{figure}

\begin{figure}
\begin{center}
\includegraphics[height=1.5in]{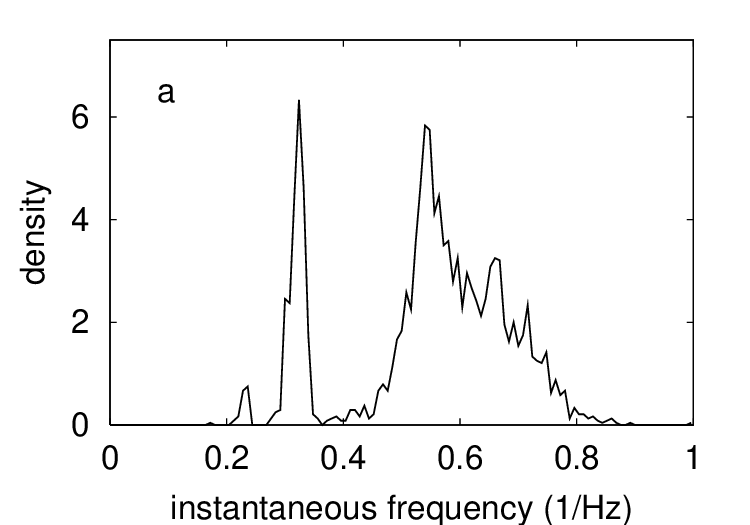}
\includegraphics[height=1.5in]{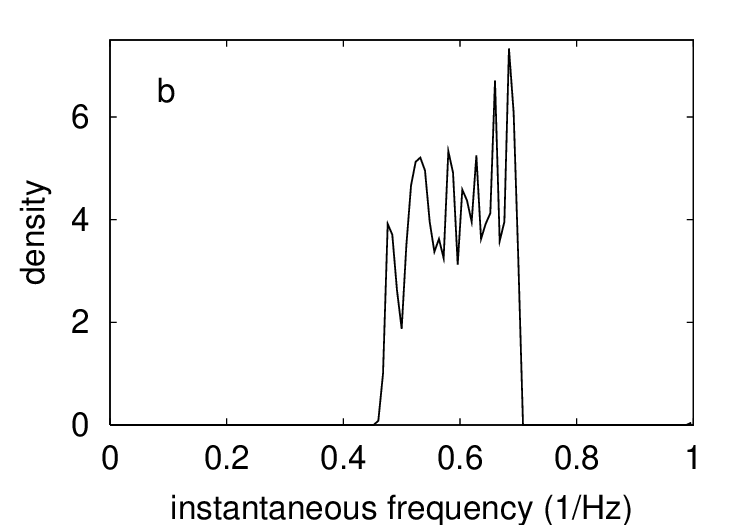}
\includegraphics[height=1.5in]{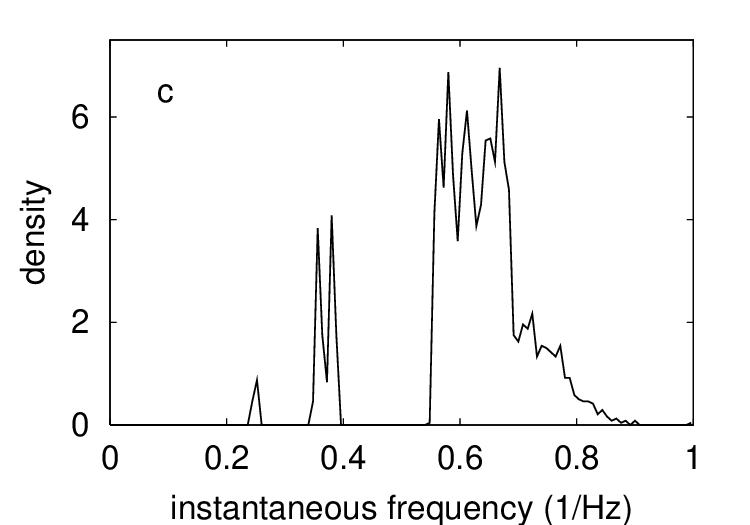}
\caption{The distributions of the instantaneous firing frequency when the
R\"{o}ssler input is applied. We set $\alpha=0.05$. (a) $a=1.88$, (b)
$a=2.94$, and (c) $a=3.31$.}
\label{fig:freq_pdf_r}
\end{center}
\end{figure}

The values of $NPE(1)$ with chaotic inputs are shown in
\FIG\ref{fig:r} for three values of $\alpha$: $\alpha=0.05$, 0.03, and
0.02. We have set $d=4$, $l_0=12$, and varied $a$. In total, 3000
interspike intervals are recorded, and the first 90\% of the data is
used for the prediction of the last 10\%. As shown in
\FIG\ref{fig:r}(a), the interspike interval coding is successful only
around $a=3.0$ for $\alpha=0.05$. The dominant frequency of $S(t)$,
which is defined as the inverse of the averaged peak intervals of
$S(t)$, coincides with the mean firing rates only within this range of
the input frequency. The firing frequency roughly falls in the range
of the internal frequency of the FHN neuron that is calculated by
applying constant bias with various amplitudes
(\FIG\ref{fig:r}(a)). The variation of the firing frequency with
chaotic inputs is larger than that with constant inputs because it is
easier to modulate interspike intervals with nonconstant inputs,
although the amount of change is limited.  Consequently, we conclude
that the neuron passes the signal information only when the input
frequency is resonant with the inherent frequency of the FHN neuron.
At the same time, interspike intervals can change to some extent to
express the temporal profile of $S(t)$ and result in the small
prediction errors.  This is guaranteed by
\FIG\ref{fig:freq_pdf_r}(b) that shows the distribution of
the instantaneous firing rates, namely, the inverse of interspike
intervals, as well as by \FIG\ref{fig:r}(a). The spike trains
carrying the filtered information on $S(t)$ can be used by downstream
neurons.  When the dominant frequency of $S(t)$ is out of the optimal
range, however, the dynamics of the neuron cannot catch up with the
temporal change of $S(t)$ that is nonresonant with the internal
frequency.  Then interspike intervals are mostly determined by
the neuronal property rather than by the inputs. Instantaneous firing
rates with the chaotic inputs are close to the inherent frequency
of the neuron or its division by 2 or 3, as shown in
\FIG\ref{fig:freq_pdf_r}(a,c). As a result, the inputs for this
range of $a$ are not encoded effectively as large values of $NPE(1)$
in \FIG\ref{fig:r}(a) indicate.

The corresponding results for $\alpha=0.03$ and $\alpha=0.02$ shown in
\FIG~\ref{fig:r}(b,c) indicate that the results above hold also for
neurons with different inherent frequencies. The range of the input
frequency associated with successful interspike-interval coding shifts
because of changes in the inherent frequency of neurons.

We also plot in \FIG~\ref{fig:r} the prediction error for neurons with
the Gaussian white noise added to $v$. The noise with standard
deviation 0.0008 is applied at each time step $dt=0.0004$. Since the
mean interspike interval is about $t=1.7$ ($\alpha=0.05$), the
standard deviation of the accumulated noise after an interspike
interval is estimated to be $0.0008 \cdot \sqrt{1.7/0.0004} = 0.052$
if we ignore the membrane leak.  The dynamical noise would typically
contribute to the change in $v$ by this amount for an interspike
interval. Figure~\ref{fig:r} shows the robustness of the results
against this amount of additive noise.

\begin{figure}
\begin{center}
\includegraphics[height=1.5in]{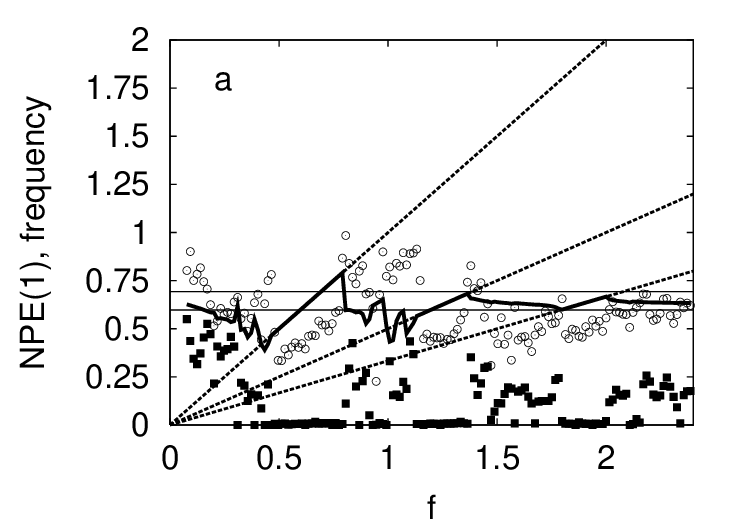}
\includegraphics[height=1.5in]{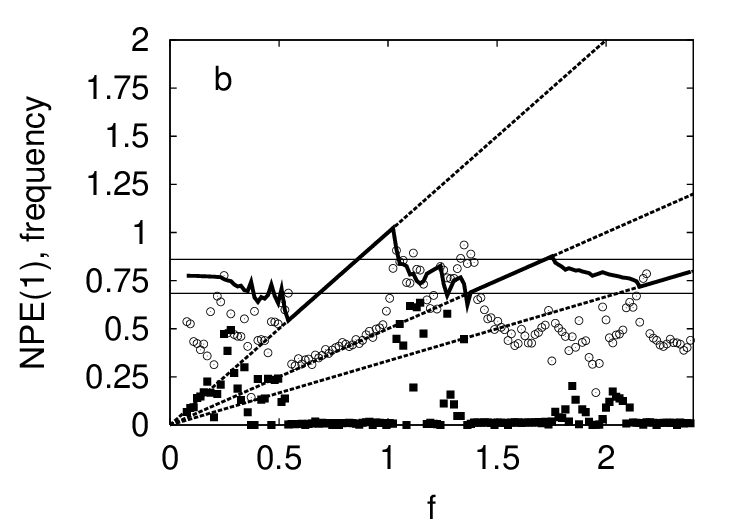}
\includegraphics[height=1.5in]{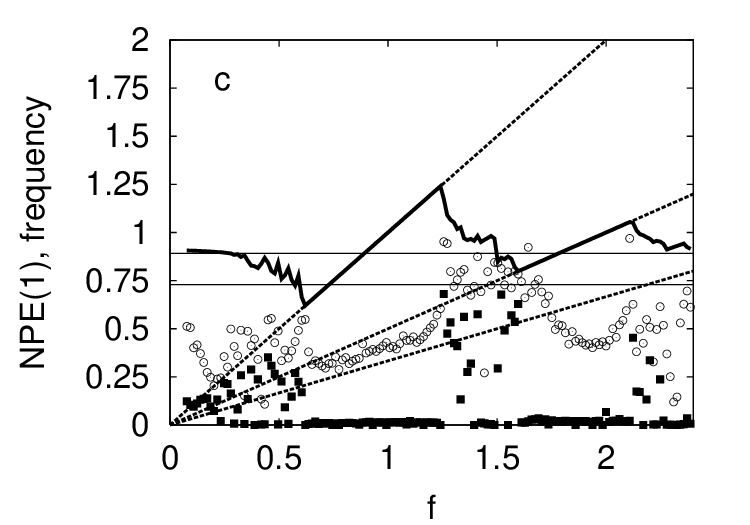}
\caption{(a) Performance of interspike interval coding in the
noiseless case (solid squares) and the noisy case (open circles).  The
neuron is driven by the quasi-periodic inputs. The frequency of inputs and
its division by 2 and 3 (dotted lines), the actual firing frequency
(thick solid line), and the inherent firing frequency range of the neurons
(thin solid lines) are also shown. (a) $\alpha=0.05$, (b)
$\alpha=0.03$, and (c) $\alpha=0.02$.}
\label{fig:s}
\end{center}
\end{figure}

The results above are qualitatively the same for nonchaotic inputs as
far as the input contains a dominant mode of oscillation. To show
this, we apply the quasi-periodic input given by
\begin{equation}
S(t) = 0.24 + 0.015 \sin (2\pi \cdot 0.2 t) + 0.04  \sin ( 2\pi f t ).
\end{equation}
The change rate of the inputs is determined by $f$ unless $f$ is
excessively small. The second term prescribes the
information signal. With this
nonchaotic case included, $NPE(h)$ measures how accurately an
interspike interval encodes the instantaneous value of $1/S(t)$
\cite{Sauer94,Racicot,Masuda_ISI}. The prediction results
with $d=4$ for three values of $\alpha$ are shown in \FIG\ref{fig:s}
in the noiseless and noisy cases as before. In \FIG\ref{fig:s}(a), for
example, small values of $NPE(1)$ are realized 
robustly with respect to $f$ in some ranges.
Around $f=0.6$, the internal and
external frequencies are in resonance.  Accordingly, the information
on the slow component of $S(t)$ superimposed on the carrier wave is
transmitted through the FHN neuron.  For $f$ around 1.2 (or 1.8), the input
frequency is twice (or three times) 
as high as the internal frequency, and the coding
is also successful in these cases.
The effect of higher resonances,
which does not appear with the R\"{o}ssler inputs, is manifested 
probably because
the superimposed sinusoidal inputs are more regular and
facilitate locking. When $S(t)$
and the neuronal dynamics are nonresonant, $NPE(1)$ is larger except
for some exceptional values of $a$ in which frequency locking with
higher harmonics is achieved.  Figure~\ref{fig:s}(b,c) corresponding
to $\alpha=0.03$ and $\alpha=0.02$ shows that this coding
feature is also observed for neurons with different inherent frequencies.
The optimal input frequency shifts accordingly.

We have shown that FHN neurons transmit the information on
chaotic and quasi-periodic inputs only when the main input frequency and the
inherent frequency of the neurons are resonant with each other. The
class II property in the exact sense, namely, independence of the
firing frequency to the inputs, enables the neurons to filter the
inputs. Filtering effects dependent on the input frequency are
actually found in experiments \cite{Gutfreund,Hutcheon00} and
related to the selection of the coding schemes, for example, in the
auditory systems \cite{Lu01}. On the other hand, the class I property
serves to express the input information in the form of interspike
intervals or instantaneous firing rates. In many realistic cases, the
neurons generally classified into class II also maintain, to some
extent, the class I property in the meaning of responsiveness of the
firing frequency to the input strength. Otherwise, the class II
neurons were too stiff to encode more than one bit of information just
by firing or nonfiring.  These properties compete with
each other to result in the tradeoff between the precision as a filter
and the performance of information expression at the output side.
Complex information processing is realized as both mechanisms are in
operation for realistic class II neurons.  They are AM (amplitude
modulation) processors in
which information on the inputs 
is encoded in the amplitude profile of the signal.
The oscillatory part composes a carrier whose dominant frequency
determines in which frequency range 
neuronal filters allow the signal to pass through.
In experiments, amplitudes of the gamma oscillation are enhanced during
electrical stimulations \cite{Traub96} and during positive theta
activities reflecting animals' exploration and sniffing
\cite{Bragin}. We postulate that the gamma oscillation in this case
specifies the family of neurons that have the
appropriate inherent frequency and allow transmission 
of signals carrying the information on behavioral or cognitive situations.

In neural networks, downstream neurons may receive spike trains from
many class II neurons for further information processing. When the
stimulus frequency and the inherent frequency of the neuron match,
a downstream neuron receives 
the superimposed incident spike trains as spike packets whose
arrivals are almost locked to the peaks of $S(t)$. In addition, the
inter-packet intervals can be modulated to some extent, which serves to
representing the input information. A downstream class II neuron
passes these information-rich quasi-regular oscillatory inputs
only when the input frequency agrees with
the inherent neuronal frequency. In this way,
a class II neuron or its cascade may receive
and filter oscillatory inputs with varying amplitudes,
whereas class I neurons as coupled
integrators or coincidence detectors may use them for other functional
purposes such as information representation by the synfire chain
\cite{Abeles91,Diesmann} and switching between the rate code and the
synchronous code \cite{Masuda_IEI,Masuda_DUAL,Vanrossum}.

\section*{Acknowledgements}

This work is
supported by the Japan Society for the Promotion of Science
and the Advanced and
Innovational Research Program in Life Sciences from the Ministry of
Education, Culture, Sports, Science, and Technology, the Japanese
Government.

\end{document}